\pdfoutput=1

\documentclass[11pt]{article}

\usepackage[]{acl}

\usepackage{times}
\usepackage{latexsym}

\usepackage[T1]{fontenc}

\usepackage[utf8]{inputenc}

\usepackage{microtype}

\usepackage{inconsolata}

\usepackage{graphicx}

\usepackage[utf8]{inputenc} 
\usepackage[T1]{fontenc}    
\usepackage{hyperref}       
\usepackage{url}            
\usepackage{booktabs}       
\usepackage{amsfonts}       
\usepackage{graphicx}       
\usepackage{nicefrac}       
\usepackage{microtype}      
\usepackage{xcolor}         
\usepackage{amsmath}
\usepackage{listings}
\usepackage{enumitem}
\setlist[itemize]{leftmargin=*}
\usepackage{multirow}
\usepackage{pifont}
\usepackage{array} 
\usepackage{ltablex} 
\usepackage{longtable} 
\keepXColumns         
\newcolumntype{Y}{>{\raggedright\arraybackslash}X}
\newcommand{\cmark}{\ding{51}}
\newcommand{\xmark}{\ding{55}}

\definecolor{codebg}{gray}{0.95}  

\title{LLMs as World Models: Data-Driven and Human-Centered Pre-Event Simulation for Disaster Impact Assessment}

\author{
Lingyao Li$^{1}$\thanks{Equal contribution as first authors.}\thanks{Corresponding authors.}, \quad
Dawei Li$^{2}$\footnotemark[1]\footnotemark[2], \quad
Zhenhui Ou$^{2}$, \quad
Xiaoran Xu$^{1}$, \\
\textbf{Jingxiao Liu$^{3}$, \quad
Zihui Ma$^{4}$, \quad
Runlong Yu$^{5}$, \quad
Min Deng$^{6}$}\\
$^{1}$ University of South Florida \quad
$^{2}$ Arizona State University \quad
$^{3}$ Massachusetts Institute of Technology\\
$^{4}$ New York University \quad
$^{5}$ University of Alabama \quad
$^{6}$ Texas Tech University\\
\texttt{lingyao@usf.edu}, 
\texttt{daweili5@asu.edu}, 
\texttt{zhenhuio@asu.edu}, 
\texttt{xiaoranxu@usf.edu},\\
\texttt{jingxiao@mit.edu}, 
\texttt{zihuima@nyu.edu}, 
\texttt{ryu5@ua.edu}, 
\texttt{mindeng@ttu.edu}
}

\begin{document}

\maketitle

\begin{abstract}
  Efficient simulation is essential for enhancing proactive preparedness for sudden-onset disasters such as earthquakes. Recent advancements in large language models (LLMs) as world models show promise in simulating complex scenarios. This study examines multiple LLMs to proactively estimate perceived earthquake impacts. Leveraging multimodal datasets including geospatial, socioeconomic, building, and street-level imagery data, our framework generates Modified Mercalli Intensity (MMI) predictions at zip code and county scales. Evaluations on the 2014 Napa and 2019 Ridgecrest earthquakes using USGS ``Did You Feel It? (DYFI)'' reports demonstrate significant alignment, as evidenced by high correlation of 0.88 and low RMSE of 0.77 as compared to real reports at the zip code level. Techniques such as RAG and ICL can improve simulation performance, while visual inputs notably enhance accuracy compared to structured numerical data alone. These findings show the promise of LLMs in simulating disaster impacts that can help strengthen pre-event planning.
\end{abstract}

\section{Introduction}

Natural disasters often disrupt infrastructure, causing significant human and economic losses \citep{jones2022human}. Efficient impact assessment is critical for emergency response and evaluating community resilience \cite{ma2024surveying}. However, most existing methods are designed for post-event  assessment, including expert inspections, ground sensors, and remote sensing~\citep{li2021social, kucharczyk2021remote, sarkar2023sam}. While effective for characterizing observed damage, these approaches are reactive by nature and offer limited utility in pre-event planning, especially for sudden-onset events like earthquakes, where early awareness is crucial~\citep{li2023exploring}. Traditional methods for pre-event simulation like scenario-based planning are useful \cite{ma2024surveying, deierlein2020cloud}, but they need extensive domain expertise for region-specific models and often lack empirical validation by addressing human-centered factors. 


Advances in large language models (LLMs) have shown promise in contextual simulation and complex reasoning across various domains \citep{li2024logicity, wang2024twin}. They are increasingly being viewed as potential world models---systems capable of learning to simulate and predict real-world scenarios \cite{wong2023word, hao2023reasoning}. Through training on large-scale datasets that encode spatial, temporal, and causal relationships, LLMs have shown potential in learning representations of how the world works. For example, current research has demonstrated their ability to understand environment status \cite{hao2023reasoning}, plan household activities \cite{xiang2023language}, and predict time-series events \cite{lee2025timecap}.

In disaster management, while LLMs are not yet widely applied as ``world models,'' researchers have explored their utility in tasks such as damage detection from satellite imagery \citep{zhang2024good} or social media \cite{wang2024near}, and emergency identification \citep{otal2024llm}. However, key research gaps still remain. First, existing studies mainly use LLMs to analyze available textual or visual data for post-event assessment, not to simulate pre-event situations. Second, while LLMs have well-demonstrated reasoning abilities, effective disaster assessment requires integrating domain-specific knowledge and data fusion so that these models can accurately reason about potential disaster scenarios. To address these limitations, this study poses the fundamental question in the context of sudden-onset disasters: \textbf{Can LLMs simulate how humans perceive seismic risks before an event occurs?} 

\begin{figure*}[htbp]
    \centering
    \includegraphics[width=1\textwidth]{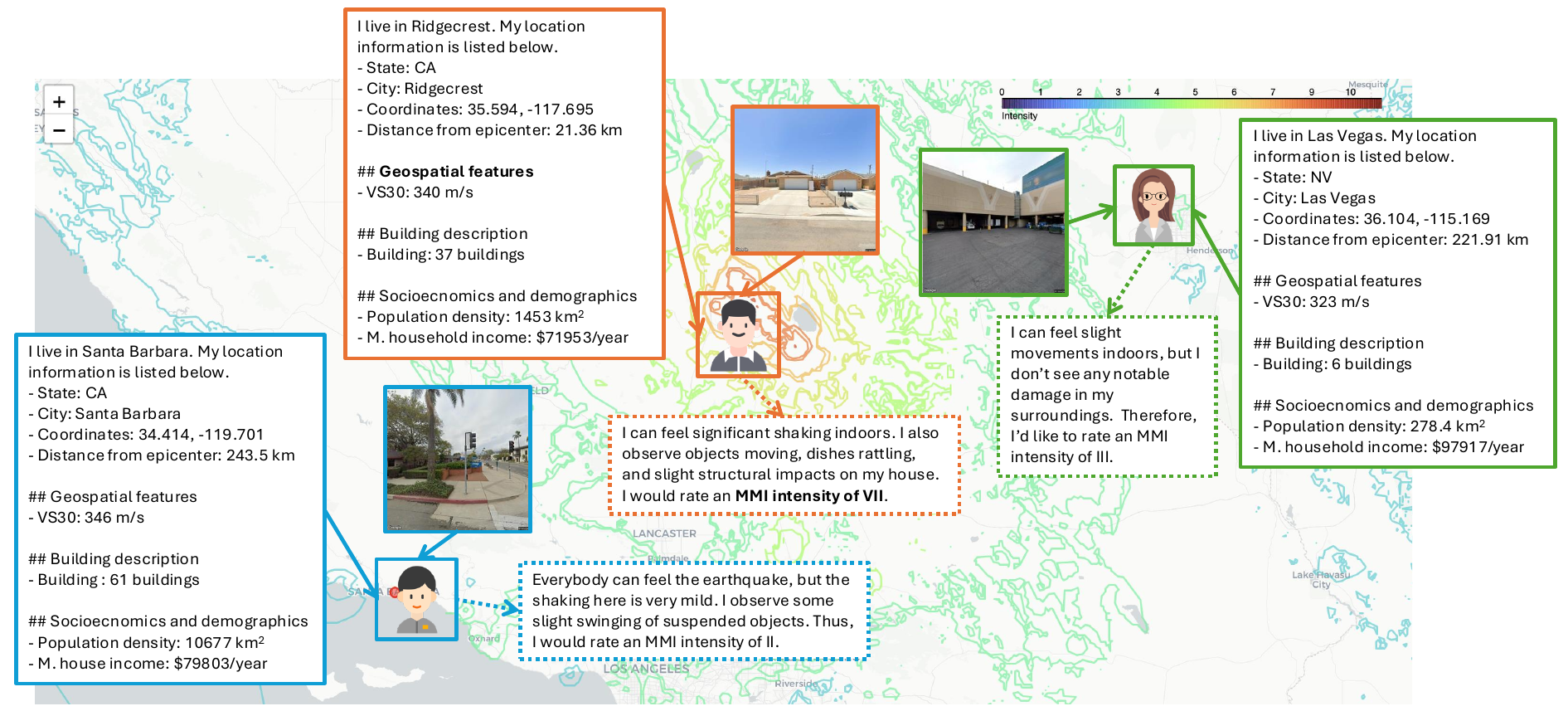}
    \vspace{-1.5em}
    \caption{An illustration of LLM-simulated human-centric sensors.}
    \label{fig:illustration}
\end{figure*}

To answer this, we develop an LLM-based framework to simulate how humans perceive seismic risks, as illustrated in Figure~\ref{fig:illustration}. By integrating rich pre-event contextual information, the LLMs are tasked with ``reasoning'' the likely severity of damage across spatial scales. Importantly, our study moves beyond theoretical simulations, grounding model evaluations in real-world events by testing on two actual earthquakes and comparing outputs against authoritative USGS reports. Our simulation demonstrates strong alignment with real reports at the zip code level, highlighting the potential of leveraging LLMs to improve pre-event planning.


\section{Related Work}
\subsection{Seismic Hazard Simulation}

Researchers have employed empirical, physics-based, and data-driven methods for simulating seismic disasters. Traditional approaches characterize earthquakes, such as the moment magnitude $M_w$ \cite{moschetti20242023}, and then utilize empirical ground-motion prediction equations (GMPEs) \cite{moschetti20242023,iervolino2023implications} or physics-based simulations \cite{deierlein2020cloud} to estimate site-specific shaking. While these simulations can capture complex local effects and rupture dynamics, they demand extensive data and computational resources. Recent data-driven methods have used machine learning to learn damage patterns from historical events, simulations, and remote sensing data \cite{CARDELLICCHIO20231956,yu2020rapid}). These AI-driven techniques offer advantages in scalability and flexible feature integration but are dependent on high-quality labeled data, may struggle with generalization, and often present challenges in interpretability.



A major gap across these models, whether physics-based or data-driven, is the limited integration of human-perceived shaking into predictive frameworks. The USGS has developed the ``Did You Feel It'' (DYFI) system~\cite{10.1785/gssrl.78.3.362}, a crowdsourced platform where individuals report the shaking intensity they experience after an earthquake. These reports are aggregated and converted into Modified Mercalli Intensity (MMI) \cite{usgs_mmi_scale}, which provide a human-centric, ground-truth reference for how seismic shaking is felt. While DYFI has been widely used to validate models or interpolate intensities after earthquakes, simulating human-centric perceived risks for pre-event could be important for developing predictive frameworks that anticipate not only physical shaking but also tangible impacts on communities.

\subsection{LLM as World Models}
Recent advance in LLMs such as GPT-4~\cite{achiam2023gpt} and deepseek-R1~\cite{guo2025deepseek} has motivated researchers to leverage LLMs to solve complex tasks, such as reasoning~\citep{li2025system,yu2025chain}  and domain-specific exploration~\citep{zhang2024shifcon,yang2024llm,zhang2024balancing}. In the context of disaster, researchers have leveraged LLMs to process multimodal information for vulnerability evaluation~\citep{martelo2024towards}, impact assessment~\citep{taghian2023disaster,akinboyewa2024automated}, information coordination~\citep{yu2024multimodal}, and recovery planning~\citep{white2025small}.

Beyond conventional question-answering and analytical tasks~\cite{yang2025quantifying}, LLMs are being used as world models~\citep{hao2023reasoning,zhao2023large} for complex simulation. While definitions of world models vary, their central concept involves leveraging forward reasoning to predict future states and dynamics in real-world scenarios. For example, they can produce large-scale simulations---an LLM-driven city simulation platform, OpenCity, runs tens of thousands of agents to emulate daily urban activities, successfully reproducing emergent patterns like mobility flows and residential segregation ~\citep{yan2024opencity}. In disaster scenarios, LLM-based world models have been explored in flooding forecasting~\cite{wang2025remflow} and evacuation planning~\cite{hostetter2024large}.


Building on these advances, we propose leveraging LLMs as simulation tools to estimate how humans might perceive and report seismic risks before an event occurs. Our work addresses two critical gaps: first, the limited pre-event simulation of seismic risk; and second, the underexplored application of LLMs as world models for deriving human-centric insights.


\section{Data and Methods}

\subsection{Framework Design}

To simulate disaster impacts before events, we propose a novel framework that treats LLM as synthetic observers. This framework conceptualizes LLMs as ``virtual sensors'' capable of ``perceiving'' multimodal inputs and ``reasoning'' about disaster risks based on the MMI levels (see appendix \ref{sec:mmi}) that approximate human perception of shaking. As illustrated in Figure~\ref{fig:illustration}, we associate each sampled spatial location with a bundle of features, including seismic distance, site conditions, local building characteristics, and socioeconomic factors. In addition, Google Street View provides a first-person view of the built environment. Collectively, this feature set closely replicates the perceptual input available to disaster responders during an event.

Formally, let each sample \(i\) be associated with a fused feature representation. We specifically select the following features that prior research has illustrated as critical predictors of seismic risk. \cite{frigerio2016gis, kassem2020seismic, riedel2015seismic, mori2020new}.

\vspace{-2mm}
\[
\mathcal{X}_i = \{ E_i, G_i, L_i, B_i, S_i, V_i \}
\]
\vspace{-6mm}

\begin{itemize}
    \item \( E_i \): Earthquake parameters (e.g., magnitude, epicenter distance, depth),
    \vspace{-3mm}
    \item \( G_i \): Geospatial features (e.g., VS30),
    \vspace{-3mm}
    \item \( L_i \): Location metadata (e.g., state, city, zip code),
    \vspace{-8mm}
    \item \( B_i \): Building attributes (e.g., number, type, height, material),
    \vspace{-3mm}
    \item \( S_i \): Socioeconomic indicators (e.g., population density, income),
    \vspace{-3mm}
    \item \( V_i \): Street-level view (Google Street image).
    \vspace{-1mm}
\end{itemize}

The LLM acts as a reasoning function \( f_\theta \) with parameters \( \theta \), generating both a reasoning trace and an MMI rating:

\vspace{-6mm}
\[
\hat{y}_i, e_i = f_\theta(\mathcal{X}_i), \quad \hat{y}_i \in \{ \text{I}, \text{II}, \ldots, \text{XII} \}
\]
\vspace{-6mm}

\begin{figure*}[htbp]
    \centering
    \includegraphics[width=1\textwidth]{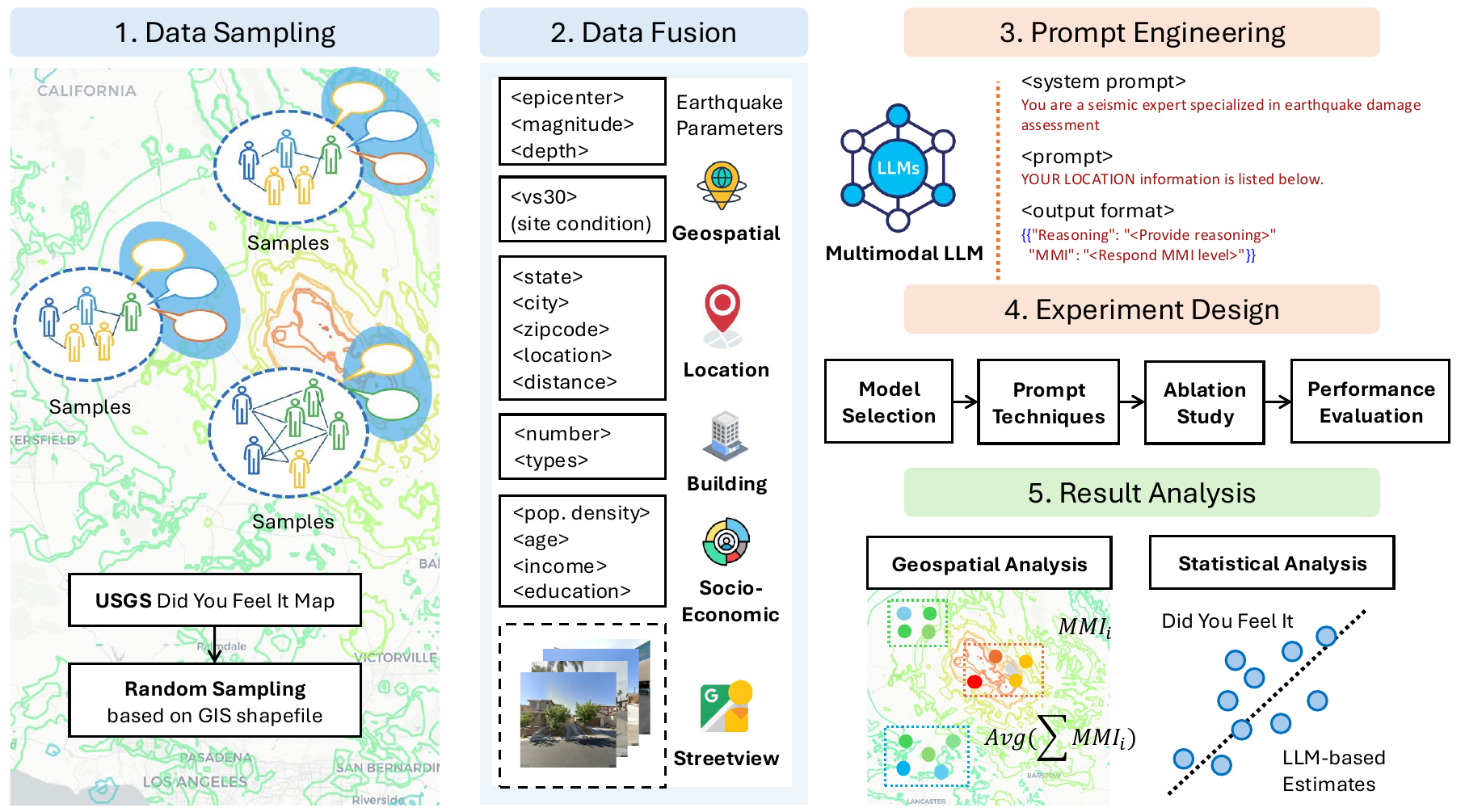}
    \vspace{-1.5em}
    \caption{An illustration of the framework design.}
    \label{fig:experiment_design}
\end{figure*}

The full pipeline consists of five components, as shown in Figure~\ref{fig:experiment_design}: (1) spatial sampling, (2) data fusion, (3) prompt engineering, (4) experiment design, and (5) result analysis, which we specifically explain in the following sections. 


\subsection{Data Sampling}

In Step 1 (Figure \ref{fig:experiment_design}), we use a polygon-based GIS shapefile to define administrative zones (e.g., zip codes) and apply within-polygon stratified random sampling to ensure spatial representativeness across the study area. Let: \(\mathcal{Z} = \{ z_1, z_2, \ldots, z_M \}\) represent the set of all zip code polygons. For each polygon \( z_j \), we sample:

\vspace{-6mm}
\[
\mathcal{P}_j = \{ p_{j1}, p_{j2}, \ldots, p_{jn_j} \} \sim \text{Uniform}(z_j)
\]
\vspace{-6mm}

ensuring that points are uniformly drawn from within each polygon’s spatial boundary. We then sample 50 data points per zip code. This stratified random sampling strategy can help ensure spatial representativeness and mitigate biases associated with population density or urban–rural areas.

\subsection{Data Fusion}

In Step 2 (Figure \ref{fig:experiment_design}), for each sampled point \( p_{ji} \), we collect and assemble the feature set \( \mathcal{X}_{ji} \) from open-source and public datasets including USGS earthquake data, OpenStreetMap building data, American Community Survey (ACS) data, and Google Maps street views.

\vspace{0.15cm}
\noindent\textbf{Earthquake data, site conditions, and location information.} We source earthquake parameters, including moment magnitude, epicentral coordinates, and focal depth, from the USGS ShakeMap product~\cite{wald2006shakemap} as $E$. To account for local site amplification and deamplification, we incorporate the USGS VS30 dataset~\cite{mcphillips2020updated} as the geospatial features $G$, a widely used proxy for near‐surface geological effects. 

We also incorporate location information, including latitude, longitude, state, county, zip code, and the distance from the epicenter for each sampled point. Additionally, we utilize MMI measurements from the USGS DYFI program as ground-truth labels, represented by $y_j$ for each zip code.

\vspace{0.15cm}
\noindent\textbf{Building description.} We use OpenStreetMap ~\cite{ramm2010openstreetmap} (OSM) data to extract building features $B$, including the total number of buildings, type distribution, height range, and average height within a 100-meter radius of each sampled point. Meanwhile, we summarize the prevalence of major construction materials (e.g., concrete, masonry, timber, steel). These features can help characterize the distribution, physical attributes, and seismic-design status of buildings within the sensor's surroundings.

\vspace{0.15cm}
\noindent\textbf{Socioeconomic and demographic factors.} We collect socioeconomic and demographic factors $S$ at the Census Block Group (CBG) level from the American Community Survey (ACS) \cite{USCensusBureauACS2022}. We spatially join the coordinates of each randomly selected sample to their corresponding CBG polygon and extract relevant ACS key indicators, including population, population density, urbanization ratio, 65- and over-age proportion, median household income, and higher education attainment rate. These variables imply a quantitative evaluation of population vulnerability in a disaster event.


\vspace{0.15cm}
\noindent\textbf{Google Maps street imagery}. We further incorporate Google Street View imagery $V$ to enrich the environmental context at each sampled location. These street-level images capture fine-grained visual cues of the surrounding built environment---such as building, vegetation, curb conditions, and street density---that are difficult to numerically encode but essential for human-like visual reasoning. Using the Google Maps API~\cite{googleMapsAPI}, we query the available image for each point in our sampling grid. This visual context allows the LLM to ``see'' the landscape as if it conducts a field visit. 

\subsection{Prompt Design}

To guide the reasoning process of the LLM and ensure interpretability and consistency, we design a prompt template that mirrors the workflow of a seismic domain expert. The prompt follows a role-based instruction format in which the model is assigned the role of a seismic specialist responsible for evaluating earthquake damage using the MMI scale. The MMI scale provides a human-centric interpretation that consists of twelve levels describing the severity of earthquake shaking. The detailed descriptors of MMI are attached in Appendix \ref{sec:mmi}. 


The prompt includes six distinct sections corresponding to the multimodal feature set \(\mathcal{X}_i\) introduced earlier: earthquake parameters \(E_i\), geospatial features \(G_i\), location metadata \(L_i\), building descriptions \(B_i\), socioeconomic context \(S_i\), and street-level view \(V_i\). The model’s response is constrained to a JSON format with two fields: a free-text reasoning explanation and the predicted MMI category (see prompt template in Appendix \ref{sec:prompt}). We implement Chain-of-Thought (CoT) to instruct LLM output a detailed reasoning justifying its assessment before final MMI estimate.




\subsection{Experiment Design}

\textbf{LLM Selection:} We select both open- and closed-source LLMs to conduct the simulation. For open-source LLMs, we use models from Llama-3.2 and Qwen-2.5 family with various sizes, as listed in Table~\ref{tab:main}.


\vspace{0.2cm}
\noindent\textbf{Prompting Techniques:} We incorporate the following prompting techniques in our experiment:
\begin{itemize}
    \item In-Context Learning (ICL): This helps LLM adapt to tasks by incorporating examples directly within the input prompt.  ICL is applied by embedding a detailed MMI reference guide directly within the prompt.
    \vspace{-2mm}
    \item Retrieval-Augmented Generation (RAG): It combines information retrieval with text generation that allows LLMs to access external data sources at inference. We provide LLM with a set of multimodal features and the reported MMI within the prompt as the retrieved context to ground their MMI estimates.
    \vspace{-3mm}
\end{itemize}



\vspace{0.2cm}
\noindent\textbf{Evaluation:}
The LLM generates a MMI prediction, $\hat{y}_{ji} = f_\theta(\mathcal{X}_{ji})$, for each sampled location $i$ within a predefined administrative area $j$ (e.g., zip code, county). These point-level predictions are then aggregated to compute an average predicted MMI for each area $j$. If area $j$ contains $n_j$ samples, its average predicted MMI $\overline{\hat{y}}_j$ is calculated as:

\vspace{-2mm}
\[
\overline{\hat{y}}_j = \frac{1}{n_j} \sum_{i=1}^{n_j} \hat{y}_{ji}
\]
\vspace{-2mm}

Similarly, a corresponding ground-truth MMI value for area $j$, denoted as $\overline{y}_j$, is derived from USGS DYFI reports. To quantify the model's predictive accuracy using these aggregated area-level values, we compute two metrics. First, the Root Mean Square Error (RMSE) is used:

\vspace{-2mm}
\[
\text{RMSE} = \sqrt{\frac{1}{N}\sum_{j=1}^{N}(\overline{\hat{y}}_j - \overline{y}_j)^2}
\]
\vspace{-2mm}

where $N$ is the total number of administrative areas being evaluated (indexed by $j$). Second, we calculate Pearson's correlation coefficients $r$ to assess the strength and direction of the association between the LLM-predicted MMI ($\overline{\hat{y}}_j$) and the ground-truth MMI ($\overline{y}_j$). These evaluations are performed at both zip code and county levels.

\begin{table*}[h]
\small
\centering
\scalebox{0.85}{
\begin{tabular}{lccccccccc}
\toprule[1.2pt]
\multirow{2}{*}{\textbf{Model}} & \multirow{2}{*}{\begin{tabular}[c]{@{}c@{}}\textbf{Open}\\ \textbf{Source}\end{tabular}} & \multicolumn{4}{c}{\textbf{Napa}}                  & \multicolumn{4}{c}{\textbf{Ridgcrest}}             \\
                       &                                                                        & $\bf{RMSE_Z\downarrow}$ & $\bf{Corr_Z\uparrow}$ & $\bf{RMSE_C\downarrow}$ & $\bf{Corr_C\uparrow}$ & $\bf{RMSE_Z\downarrow}$ & $\bf{Corr_Z\uparrow}$ & $\bf{RMSE_C\downarrow}$ & $\bf{Corr_C\uparrow}$ \\ \hline \addlinespace[2pt]
GPT-4o-2024-08-06      & \xmark                                                  & 2.43     & \bf{0.77}     & 2.37     & \bf{0.88}     & 1.97     & \bf{0.75}     & 1.91     & \bf{0.77}     \\
GPT-4.1-mini           & \xmark                                                  & 2.56     & 0.61     & 2.48     & 0.67     & \bf{0.92}     & 0.64     & \bf{0.77}     & 0.76     \\
Claude-3.5-haiku       & \xmark                                                  & 2.11     & 0.58     & 2.05     & 0.70     & 1.35     & 0.59     & 1.38     & 0.71     \\ \hline \addlinespace[2pt]
Llama-3.2-11B-VI       & \cmark                                                  & 3.19     & 0.44     & 3.05     & 0.86     & 3.22     & 0.33     & 3.22     & 0.27     \\
Llama-3.2-90B-VI       & \cmark                                                  & 2.62     & 0.57     & 2.55     & 0.66     & 2.06     & 0.62     & 2.19     & 0.59     \\
Qwen2.5-VL-3B          & \cmark                                                  & 3.63     & 0.29     & 3.59     & 0.15     & 3.88     & 0.01     & 4.08     & -0.20    \\
Qwen2.5-VL-7B          & \cmark                                                  & 1.79     & 0.43     & 1.68     & 0.70     & 1.53     & 0.05     & 1.59     & -0.18    \\
Qwen2.5-VL-32B         & \cmark                                                  & \bf{1.59}     & 0.70     & \bf{1.56}     & 0.79     & 0.99     & 0.71     & 0.96     & 0.80     \\
Qwen2.5-VL-72B         & \cmark                                                  & 2.17     & 0.46     & 2.12     & 0.44     & 1.39     & 0.64     & 1.28     & 0.86     \\ \toprule[1.2pt]
\end{tabular}}
\caption{Main experiment results, the best result for each metric is in bold.}
\label{tab:main}
\end{table*}

\section{Experimental Results}

We select two cases to demonstrate our proposed framework: (1) the 2014 Napa earthquake (magnitude 6.0) and (2) the 2019 Ridgecrest earthquake (magnitude 7.1), both of which occurred in California, U.S. (see details in Appendix~\ref{sec:scenario}). For each case, we use the USGS DYFI reports as the ground-truth dataset \cite{usgs_napa_earthquake, usgs_ridgecrest_earthquake}. Additionally, for each case, We compile feature sets for 50 sample points from each of the top 100 zip codes with the highest number of responses, resulting in 5,000 samples per event. Due to limitations in Google image availability for the Napa case, only 4,920 samples are retrieved. Comparisons with DYFI data are first conducted at the zip code level, where each of the 100 aggregated values represents the average of 50 simulated samples.

Figure~\ref{fig:gis} presents the spatial distribution of predicted MMI at the zip code level for the Napa and Ridgecrest earthquakes. This visualization highlights variations in simulated seismic impacts across geographical areas and among different LLMs. Based on the best-performing models (lowest RMSE: GPT-4.1-mini for the 2019 Ridgecrest earthquake and Qwen-2.5-32B for the 2014 Napa earthquake), we observe consistent geospatial patterns in both cases. Specifically, the simulations indicate elevated perceived risk near the epicenter (a red star mark in Figure~\ref{fig:gis}), with diminishing simulated impact as distance increases. Moreover, the LLM-based predictions align well with DYFI reports from these two events: it is important to note that the Napa earthquake, despite its lower magnitude, led to more significant impacts.

The following sections are organized below. First, we evaluate the performance of the selected LLMs, comparing their accuracy using quantitative metrics, and examining the influence of model scaling and prompting strategies. Next, we conduct an input feature analysis to evaluate how different data modalities can impact predictive performance. Lastly, the output reasoning analysis explores the internal decision-making processes of the models, which identifies linguistic nuances that illustrate how LLMs interpret the inputs.

\subsection{Model Performance}


Before the main experiment, we perform a data leakage test using two close-source models---Claude-3.5-haiku and GPT-4.1-mini---to show that our simulation is free of data leakage issues (see Appendix~\ref{sec:leakage}). The main experiment results are shown in Table~\ref{tab:main}, from which we draw the following findings:

\begin{figure}[h]
    \centering
    \includegraphics[width=0.42\textwidth]{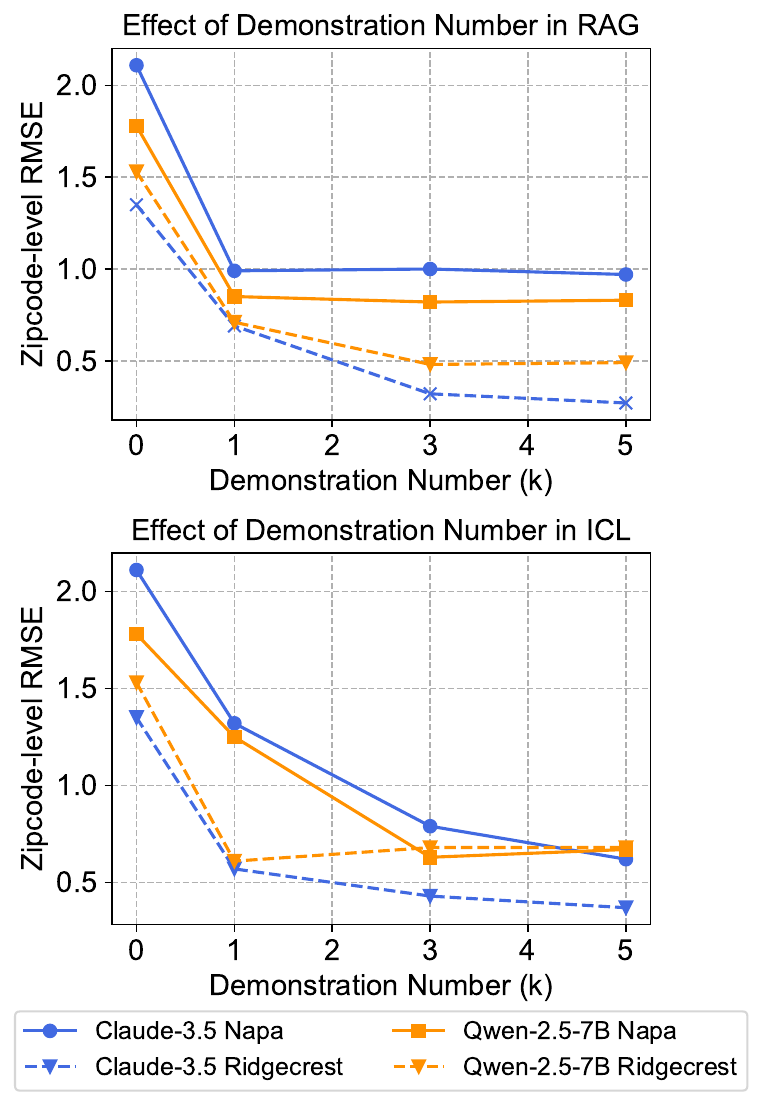}
    \vspace{-0.5em}
    \caption{Demonstration analysis on RAG and ICL.}
    \label{fig:demonstration}
\end{figure}

\noindent\textbf{LLMs deliver promising simulation results.} As shown in Table~\ref{tab:main}, all of the best correlation scores across datasets and area levels exceed 0.7, indicating a strong alignment between the predicted outcomes and the ground truth impact labels. These high correlation values suggest that LLMs hold significant potential for effectively simulating human-perceived risks in disaster scenarios.

\vspace{0.2cm}
\noindent\textbf{Closed-source LLMs generally outperform open-source models.}
Table~\ref{tab:main} shows a clear trend that closed-source LLMs consistently outperform open-source counterparts, achieving the best results in 6 out of 8 cases. This suggests that commercial closed-source models possess stronger geospatial reasoning capabilities and align more closely with human judgment in disaster sensing tasks~\cite{li2024generation}. Among the open-source models, Qwen-2.5-32B obtains the top results with lowest RMSE for the Napa case.

\vspace{0.2cm}
\noindent\textbf{RMSE and correlation can be inconsistent.}
Another noteworthy observation from Table~\ref{tab:main} is that the two evaluation metrics---correlation and RMSE---do not always align. For instance, Llama-3.2-11B shows a high correlation but a poor RMSE at the county level for the Napa case. This discrepancy arises because correlation captures the model's ability to predict relative ordering of seismic impacts, whereas RMSE reflects the absolute prediction errors. Thus, even when models effectively estimate the relevant severity (lower RMSE), they may not correctly distinguish between higher- or lower-impact areas (lower correlation). 


We then evaluate the influence of incorporating demonstration methods with RAG and ICL on the simulation, as shown in Figure~\ref{fig:demonstration}. These demonstration techniques can enhance the capacity of LLMs to generalize from provided information. In particular, the enhanced models consistently show higher performance, illustrated by decreasing $RMSE_Z$. It is also well-noted that even limited demonstrations can greatly improve model predictions.



\subsection{Input Feature Analysis}



\begin{figure}[h]
    \centering
    \includegraphics[width=0.46\textwidth]{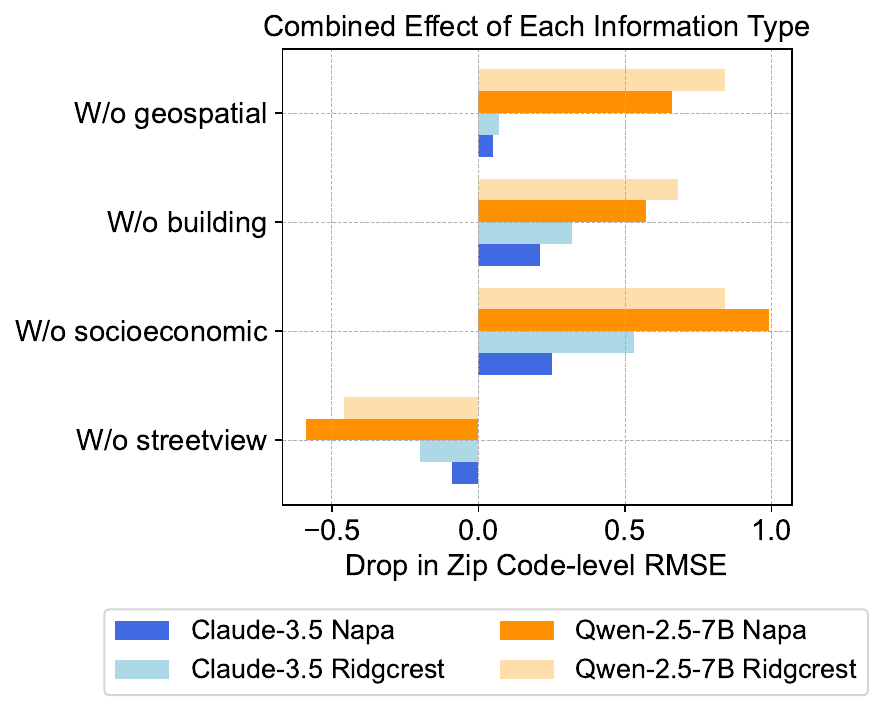}
    \caption{Input information feature analysis results.}
    \label{fig:information}
\end{figure}

In this section, we examine how input features beyond earthquake and location information affect LLMs' simulation performance. We conduct experiments using Claude-3.5-haiku and Qwen-2.5-7B, with the results presented in Figure~\ref{fig:information}. Interestingly, we find that only street view information contributes to improved simulation performance. In contrast, removing any of the other three features alone---geospatial, building, or socioeconomic data---can decrease the zip code-level RMSE.

We attribute this performance degradation to several potential factors: (1) limitations of LLMs in processing numerical information as they could complicate LLM's reasoning process~\cite{yin2024sparrow,bodensohn2025unveiling}; (2) the lack of domain-specific knowledge required to interpret geospatial, building, and community-related data~\cite{gao2024raw}; (3) the inherent limitations of the self-attention mechanism in capturing spatial adjacency and performing geometric reasoning~\cite{requeima2024llm}.


\subsection{Output Reasoning Analysis}


\begin{figure*}[h]
    \centering
    \includegraphics[width=1\textwidth]{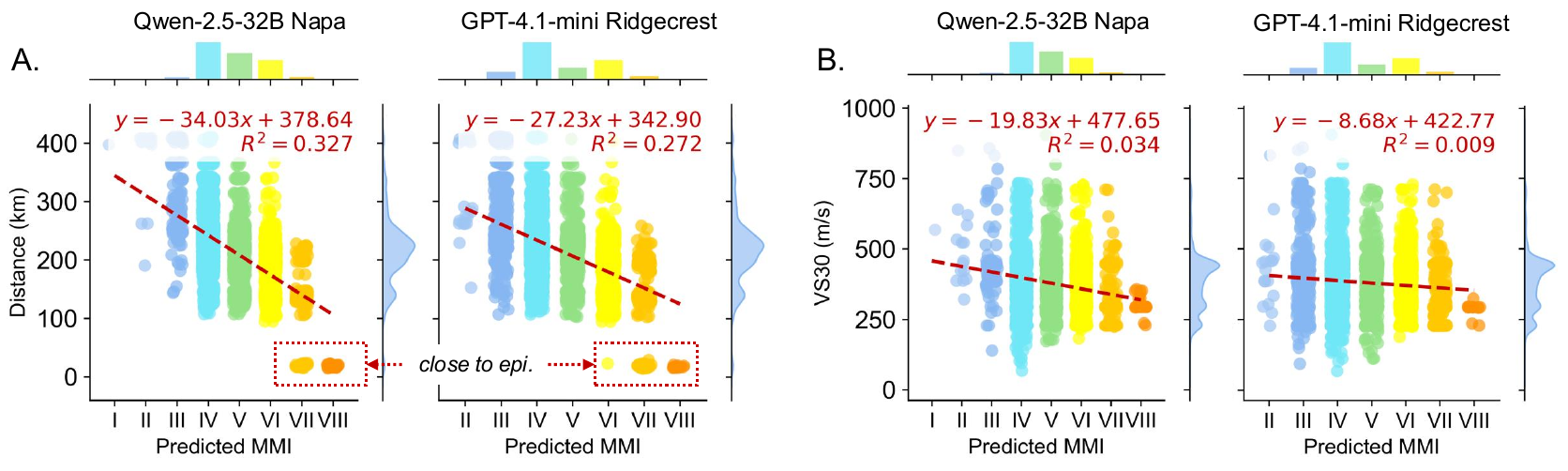}
    \vspace{-2em}
    \caption{Output reasoning analysis in terms of (A) distance (where the x-axis is the predicted MMI, and y-axis is the distance from the epicenter (km)) and (B) VS30 (where the x-axis is the predicted MMI, and the y-axis the local site condition represented by VS30 (m/s)).}
    \label{fig:result_distance}
\end{figure*}

\begin{figure*}[h]
    \centering
    \includegraphics[width=1\textwidth]{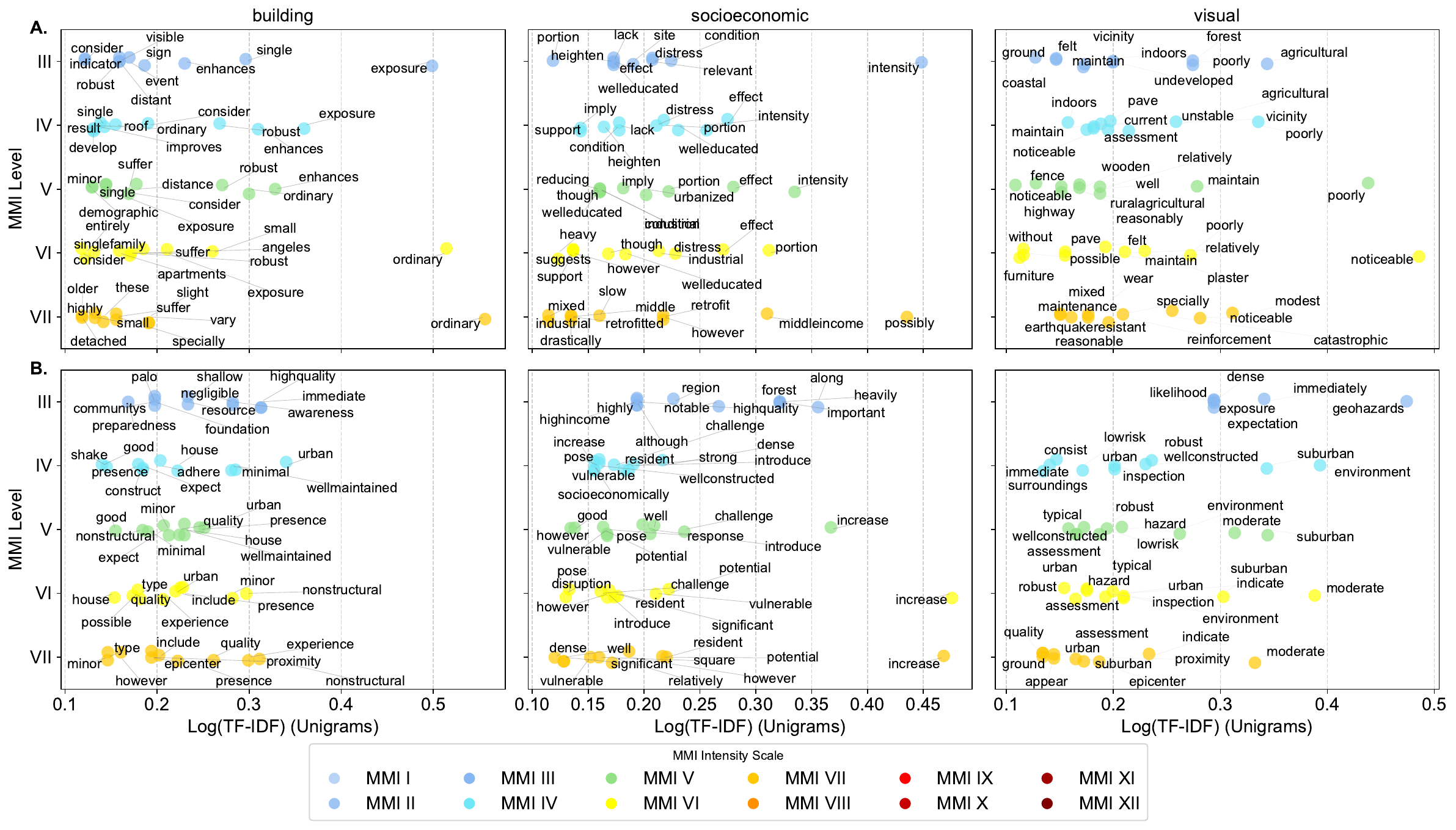}
    \vspace{-2em}
    \caption{Output reasoning analysis in terms of different input features with (A) Qwen-2.5-32B for the Napa case, and (B) GPT-4.1-mini for the Ridgecrest case. The x-axis is the log(TF-IDF), while the y-axis is the predicted MMI.}
    \label{fig:result_text}
\end{figure*}

To complement our quantitative evaluations, we examine how GPT-4.1-mini and Qwen-2.5-32B (lowest RMSE) reason when predicting MMI values. We summarize our findings as below:

\vspace{0.2cm}
\noindent\textbf{LLMs capture seismic attenuation but underutilize local site conditions.} As shown in Figure~\ref{fig:result_distance}, both models display a clear negative correlation between epicentral distance and predicted MMI, most notably in Qwen2.5-32B’s Napa earthquake predictions, which indicates that LLMs have internalized the concept of seismic attenuation. However, the relationship between VS30 values (a proxy for local ground conditions) and MMI is weak across both models. High MMI values occur almost exclusively near the epicenter, suggesting limited sensitivity to local site effects.

\vspace{0.2cm}
\noindent\textbf{LLMs use distinct lexical cues for MMI reasoning across multimodal inputs.} Figure~\ref{fig:result_text} presents a taxonomy of language used by both models across three perspectives: buildings, socioeconomic context, and street-level imagery. Unigram analysis reveals that GPT-4.1-mini and Qwen2.5-32B employ different vocabularies and reasoning styles. For the building assessment, GPT-4.1-mini adopts a descriptive and hedged reasoning style. At low MMI levels, it uses terms like ``enhance,'' ``robust,'' and ``improve,'' while shifting to ``suffer'' and ``detached'' at higher levels. Qwen2.5-32B relies on more technical terms, such as ``compliance'' and ``stringent'' at low levels, and ``crack'' and ``susceptibility'' at higher ones.

\vspace{0.2cm}
\noindent\textbf{Socioeconomic reasoning diverges in focus and tone.} GPT-4.1-mini links lower MMI levels to terms like ``urbanized'' and ``welleducated,'' and higher levels to ``industrial'' and ``heavy,'' occasionally incorporating cautious language such as ``possibly'' or ``suggest.'' Qwen2.5-32B emphasizes systemic vulnerability, shifting from ``limited'' and ``stable'' to ``cascade'' and ``amplification'' as predicted MMI increases.

\vspace{0.2cm}
\noindent\textbf{Visual reasoning contrasts environmental vs. structural emphasis.} GPT-4.1-mini references broader environmental cues, from ``forest'' and ``agricultural'' at lower levels to ``catastrophic'' at higher ones. In contrast, Qwen2.5-32B focuses on structural compliance, mentioning ``firm,'' ``code,'' and ``reinforced'' in a progression of seismic risk.

\section{Discussion}
Based on our experimental result analysis, we conclude the following insights and opportunities for future study:

\begin{itemize}
    \item LLMs can effectively simulate human-centric seismic risks, showing strong alignment with real-world USGS DYFI reports. This can help develop pre-event impact assessment. One promising direction is to explore broader application in LLM-based simulations for disaster responses.
    \vspace{-3mm}
    \item Simulation performance depends on model type, size, and input information: closed-source and larger models generally perform better; techniques like RAG and ICL help; street-level imagery boosts accuracy, while structured data may hinder it due to modality alignment limitations. Future works can explore more effective prompting strategies and reasoning structures to further improve the simulation performance.
    \vspace{-3mm}
    \item LLMs show diverse reasoning styles and strong practical value, as model-specific rhetorical patterns reflect architectural and data differences. These differences suggest the potential impact of training data and model architecture on LLMs' reasoning styles. It would be meaningful to further explore the reasoning mechanisms of LLMs when used as world models. 
\end{itemize}

\vspace{-2mm}
Our study has significant practical implications. Leveraging LLMs and open-source data to simulate seismic risk enables authorities to proactively assess potential disaster impacts. Integrating our framework into early-warning systems can also help identify vulnerable communities and enhance disaster response.

\section{Conclusions}

Our study demonstrates the potential of using LLMs to simulate seismic risk before an earthquake occurs. The alignment between model predictions and real-world reports highlights the importance of multimodal inputs and advanced LLM techniques like RAG and ICL in simulation settings. Moreover, our findings reveal that simulation performance can vary across LLMs and input features. These results make a meaningful contribution to advancing data-driven, human-centric simulation with LLMs for real-world scenarios.



\newpage
\section{Limitations}

Several limitations warrant further consideration. First, our experiments focus only on two earthquake cases---the 2014 Napa and 2019 Ridgecrest events---which may not fully represent global variations in seismic hazards, urban densities, and construction practices. Therefore, generalizability requires additional validation.

Second, despite employing stratified random sampling within zip codes, potential biases remain due to gaps in data availability, such as incomplete Google Street View coverage or inconsistencies in socioeconomic and building datasets. This may lead to underrepresentation of certain neighborhoods or misalignment with high-risk zones.

Third, although our framework integrates diverse heterogeneous data, we do not conduct detailed feature selection or examine individual parameters (e.g., housing age, infrastructure proximity). Consequently, interpreting model behavior at a granular level requires further investigation.


\section{Ethics}

\noindent\textbf{Responsible data use.} Our research relies exclusively on publicly available and anonymized datasets, including geospatial, demographic, and imagery data from USGS, OpenStreetMap, and Google Street View. All analyses are conducted at aggregated administrative levels (zip code and county), minimizing risks to individual privacy.

\vspace{0.2cm}
\noindent\textbf{Deployment guidance.} It is important to recognize that LLM-based simulations cannot fully capture the complexity and diversity of individual experiences in real disaster scenarios. Therefore, model predictions should be viewed as supportive tools rather than replacements for expert judgment, and used responsibly with validated empirical assessments in practical decision-making contexts.

\bibliography{custom}

\appendix
\onecolumn
\section{Modified Mercalli Intensity (MMI) Scale}
\label{sec:mmi}

Table~\ref{tab:mmi_scale} shows the MMI scale used to support the classification of seismic risks in our study. 


\begin{table*}[htbp]
\centering
\small
\begin{tabular}{lp{13cm}}
\toprule[1.2pt]
\textbf{MMI Level} & \textbf{Description} \\
\midrule
I & Not felt except by a very few under especially favorable conditions. \\
\hline
II & Felt only by a few persons at rest, especially on upper floors of buildings. Delicately suspended objects may swing. \\
\hline
III & Felt quite noticeably by persons indoors, especially on upper floors of buildings. Many people do not recognize it as an earthquake. Standing motor cars may rock slightly. Vibration similar to the passing of a truck. Duration estimated. \\
\hline
IV & Felt indoors by many, outdoors by few during the day. At night, some awakened. Dishes, windows, doors disturbed; walls make cracking sound. Sensation like heavy truck striking building. Standing motor cars rocked noticeably. \\
\hline
V & Felt by nearly everyone; many awakened. Some dishes, windows broken. Unstable objects overturned. Pendulum clocks may stop. \\
\hline
VI & Felt by all, many frightened. Some heavy furniture moved; a few instances of fallen plaster. Damage slight. \\
\hline
VII & Damage negligible in buildings of good design and construction; slight to moderate in well-built ordinary structures; considerable damage in poorly built or badly designed structures; some chimneys broken. \\
\hline
VIII & Damage slight in specially designed structures; considerable damage in ordinary substantial buildings with partial collapse. Damage great in poorly built structures. Fall of chimneys, factory stacks, columns, monuments, walls. Heavy furniture overturned. \\
\hline
IX & Damage considerable in specially designed structures; well-designed frame structures thrown out of plumb. Damage great in substantial buildings, with partial collapse. Buildings shifted off foundations. \\
\hline
X & Some well-built wooden structures destroyed; most masonry and frame structures destroyed with foundations. Rail bent. \\
\hline
XI & Few, if any (masonry) structures remain standing. Bridges destroyed. Rails bent greatly. \\
\hline
XII & Damage total. Lines of sight and level are distorted. Objects thrown into the air. \\
\toprule[1.2pt]
\end{tabular}
\vspace{-2mm}
\caption{Description of the Modified Mercalli Intensity (MMI) scale \cite{usgs_mmi_scale}.}
\label{tab:mmi_scale}
\end{table*}

\section{Prompt Template Design}
\label{sec:prompt}
\begin{lstlisting}[language=Python]
SYSTEM_PROMPT = """
You are a seismic expert specialized in earthquake damage assessment and disaster response. You analyze earthquake data, local conditions, and building characteristics to provide damage assessments using the Modified Mercalli Intensity (MMI) scale.
"""
\end{lstlisting}

\begin{lstlisting}[language=Python]
EARTHQUAKE_PROMPT = """
The earthquake happened date is 2025-06-01. 

Here is the EARTHQUAKE information. 
- Epicenter: {eq_place}
- Coordinates: {eq_lat}, {eq_lng}
- Magnitude: {eq_magnitude} mw
- Depth: {eq_depth} km

YOUR LOCATION information is listed below. 
- State: {state}
- City: {city}
- Zipcode: {zipcode}
- Coordinates: {lat}, {lng}
- Distance from epicenter: {distance} km

## Geospatial features in YOUR LOCATION
- VS30 at your location: {vs30} m/s 
(VS30 represents the time-averaged shear-wave velocity (VS) to a depth of 30 meters, which is a key index to account for seismic site conditions)

## Building Description in YOUR LOCATION (within a 100-meter radius)
- Building description: {building} 

## Community Socioecnomics and Demographics in YOUR LOCATION (at Cencus Block Group level)
- Population density: {population_density} people per square km
- Urban population percentage: {urban_population_pct}%
- Over 65 percentage: {over_65_rate}%
- Median household income: ${median_household_income}/year
- Education (bachelor's or higher): {education}%

## Visual Context in YOUR LOCATION
The image provided shows your surrounding environment and infrastructure. 

Based on the information provided, ASSESS the potential earthquake damage level using the Modified Mercalli Intensity (MMI) scale.
1. Identify the damage level.
2. Explain your reasoning by addressing the following factors and considering the visual context. 
   - Distance to the epicenter and earthquake magnitude
   - Geospatial features
   - Infrastructure quality and building characteristics
   - Population density and socioeconomic vulnerabilities
   - Visual image of surroundings

The following is an abbreviated description of the 12 levels of Modified Mercalli intensity. {MMI Scale}

Output the result in JSON format:
{{
    "Reasoning": "<Provide reasoning>"
    "MMI": "<Respond MMI level>",
}}
"""
\end{lstlisting}

\section{Earthquake Scenarios}
\label{sec:scenario}
\noindent\textbf{2014 Napa Earthquake} \cite{usgs_napa_executive}. On August 24, 2014, a magnitude 6.0 earthquake struck near Napa, California, causing significant structural damage despite its moderate magnitude. Approximately 613 buildings were tagged for various degrees of structural integrity concerns, including fractures, road cracks, and damage to wine storage facilities. The earthquake resulted in one death and nearly 200 injuries.

\vspace{0.2cm}
\noindent\textbf{2019 Ridgecrest Earthquake} \cite{usgs_ridgecrest_executive}. The Ridgecrest earthquake occurred on July 6, 2019, with a magnitude of 7.1, significantly larger than the Napa event but with fewer human casualties. The quake damaged around 50 homes, caused gas leaks and road cracks, and triggered fires in residential properties. Significant infrastructural damage occurred at the Naval Air Station, and widespread power outages were reported.

\section{Data Leakage Test}
\label{sec:leakage}
To further assess the potential for data leakage in the LLMs used in our experiments, we conduct a leakage test on Claude-3.5-haiku and GPT-4.1-mini, the two best-performing models in our main study. Specifically, we remove city and state names from the prompt—two elements most likely to serve as shortcuts for the models to associate with MMI levels and potentially memorize. As shown in Table~\ref{tab:data_leakage}, the removal of location information does not significantly affect the models' simulation performance. It is fair to rule out the possibility of data leakage in our main results.

\begin{table*}[h]
\centering
\small
\scalebox{1}{
\begin{tabular}{lcccccccc}
\toprule[1.2pt]
\multirow{2}{*}{\textbf{Model}} & \multicolumn{4}{c}{\textbf{Earthquake prompt}}                                  & \multicolumn{4}{c}{\textbf{\begin{tabular}[c]{@{}c@{}}Earthquake prompt\\ w/o location\end{tabular}}} \\ 
& \textbf{RMSE\_Z} & \textbf{Corr\_Z} & \textbf{RMSE\_C} & \textbf{Corr\_C} & \textbf{RMSE\_Z}       & \textbf{Corr\_Z}       & \textbf{RMSE\_C}      & \textbf{Corr\_C}      \\ \hline \addlinespace[2pt]
claude-3-5-haiku                & 2.11             & 0.58                & 2.05             & 0.70                & 2.35                   & 0.38                      & 2.26                  & 0.62                     \\
gpt-4.1-mini                    & 2.56             & 0.61                & 2.48             & 0.67                & 2.67                   & 0.62                      & 2.58                  & 0.73                    \\ \toprule[1.2pt]
\end{tabular}
}
\vspace{-2mm}
\caption{Experiment results on data leakage test.}
\label{tab:data_leakage}
\end{table*}

\begin{table}[h]
    \centering
    \small
    \begin{tabular}{lcccccc}
        \toprule[1.2pt]
        \textbf{Model Family} & \textbf{Model Size} & \textbf{Napa RMSE} & \textbf{Ridgecrest RMSE} \\
        \midrule
        \multirow{2}{*}{Llama-3.2} 
        & 11B & 3.19 & 3.22 \\
        & 90B & 2.62 & 2.06 \\
        \midrule
        \multirow{4}{*}{Qwen-2.5} 
        & 3B & 3.63 & 3.88 \\
        & 7B & 1.79 & 1.53 \\
        & 32B & 1.59 & 0.99 \\
        & 72B & 2.17 & 1.39 \\
        \bottomrule[1.2pt]
    \end{tabular}
    \vspace{-2mm}
    \caption{Scaling law analysis: Zipcode-level RMSE across model sizes (in billions of parameters).}
    \label{tab:scaling_law}
\end{table}

\section{Spatial Distribution of Predicted MMI by LLMs}
\label{sec:gis}

\begin{figure*}[h!]
    \centering
    \includegraphics[width=1\textwidth]{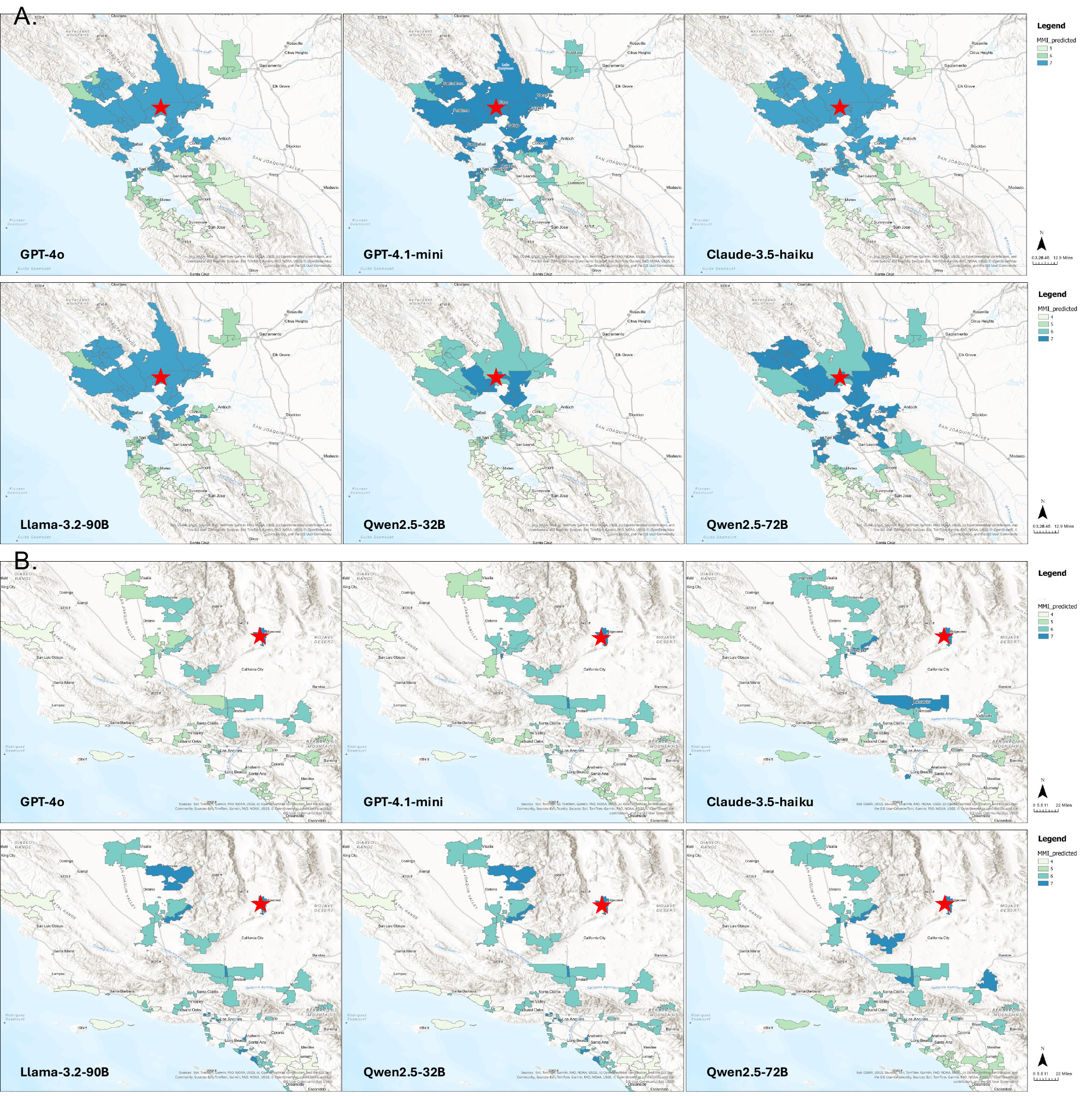}
    \vspace{-1.5em}
    \caption{Spatial distribution of predicted MMI at zip code level: (A) The 2014 Napa earthquake, and (B) the 2019 Ridgecrest earthquake based on GPT4o, GPT-4.1-mini, Claude-3.5-haiku, Llama-3.2-90B, Qwen-2.5-32B, and Qwen-2.5-72B. The red star sign shows the epicenter of the earthquake. These maps compare outputs from different LLMs, showing consistent geospatial patterns with high-intensity predictions concentrated near epicenters}
    \label{fig:gis}
\end{figure*}

Figure~\ref{fig:gis} shows the spatial distribution of predicted MMI for the 2014 Napa earthquake and the 2019 Ridgecrest earthquake at zip code level from different LLMs.

\section{Scaling Law Analysis}
To investigate the impact of LLM scaling laws on simulation performance, we conduct a scaling analysis using the Llama-3.2 and Qwen-2.5 model families. As shown in Table~\ref{tab:scaling_law}, we observe that performance in simulation generally improves with model size, excepting Qwen-2.5-72B. This indicates that the simulation performance tend to be strengthened with a larger size of LLMs.


\section{Samples of LLM Output}

\begin{table*}[h]
\centering
\small
\renewcommand{\arraystretch}{1.2}
\begin{tabular}{p{\textwidth}}
\toprule[1.2pt]
\textbf{Samples of LLM Reasoning Outputs with Different MMI Predictions} \\
\midrule
\textbf{GPT-4.1-mini | 2019 Ridgecrest | Predicted MMI: III} \\
The earthquake has a high magnitude of 7.1 Mw, which is capable of causing significant damage near the epicenter. However, the location in question, San Diego, CA, is approximately 339 km away from the epicenter in Ridgecrest, CA. At this distance, seismic waves would have attenuated considerably, reducing the intensity of shaking experienced. The VS30 value of 417 m/s indicates a site with relatively soft soil conditions, which can amplify shaking somewhat, but given the large distance, this effect is limited. The building stock consists mostly of general buildings and schools, likely built to modern California seismic codes, which enhances resilience. The population density is moderate, with a well-educated and relatively affluent community, reducing socioeconomic vulnerabilities. The visual context shows a well-maintained urban street with no visible signs of damage or distress, supporting the assessment of low shaking impact. Considering all these factors, the expected Modified Mercalli Intensity (MMI) level in San Diego would be around III, where shaking is noticeable indoors but causes no damage. \\
\hline
\textbf{GPT-4.1-mini | 2019 Ridgecrest | Predicted MMI: IV} \\
The earthquake has a high magnitude of 7.1 Mw, which is capable of causing significant damage near the epicenter. However, the location in Ventura, CA is approximately 221 km away from the epicenter in Ridgecrest, CA, which greatly reduces the intensity of shaking experienced. The VS30 value of 391 m/s indicates a site with relatively soft soil conditions, which can amplify shaking somewhat, but the distance still plays a dominant role in reducing impact. The building information is not available, but the visual context shows modern infrastructure such as a highway bridge and vehicles, suggesting reasonably good construction standards typical of California. The population density is low (65 people/km\(^2\)) with a moderate urban population percentage (25.43 \%), and socioeconomic indicators such as median income and education level are moderate, implying some resilience but also potential vulnerabilities in older or poorly maintained structures. Given the distance, moderate site conditions, and infrastructure, the expected shaking intensity would be moderate with little to no structural damage. This corresponds to an MMI level around IV to V, where shaking is felt by many but damage is minimal or limited to minor non-structural effects. \\
\hline
\textbf{GPT-4.1-mini | 2019 Ridgecrest | Predicted MMI: V} \\
The earthquake had a high magnitude of 7.1 Mw, which is capable of causing severe damage near the epicenter. However, the location in question, Apple Valley, CA, is approximately 152.25 km away from the epicenter, which significantly reduces the intensity of shaking experienced. The VS30 value of 607 m/s indicates relatively firm ground conditions, which typically reduce amplification of seismic waves. The visual context shows a rural, sparsely populated area with no visible dense infrastructure or tall buildings, suggesting limited exposure to structural damage. The population density is low (18.73 people per square km), and the urban population is only about 32.65 \%, further reducing potential damage and casualties. Socioeconomic factors such as median income and education level are moderate, but given the lack of detailed building information and the rural setting, the risk of significant structural damage is low. Considering these factors, the shaking would likely be felt by many but cause only slight damage, consistent with MMI level VI. \\
\hline
\textbf{Qwen-2.5-32B | 2014 Napa | Predicted MMI: VI} \\
The location is 19.27 km from the epicenter of a 6.0 Mw earthquake, indicating strong shaking. The VS30 value of 344 m/s suggests stiff soil, reducing amplification but not eliminating damage due to proximity. The presence of general buildings and houses, along with a rural setting and a high percentage of older residents, points to moderate to strong damage. The visual context shows a well-maintained residential area, which may indicate better-quality buildings, but this is not definitive. \\
\hline
\textbf{Qwen-2.5-32B | 2014 Napa | Predicted MMI: VII} \\
The earthquake has a magnitude of 6.0 Mw, and the location is 14.59 km from the epicenter, indicating strong shaking. The VS30 value of 354 m/s suggests moderate amplification. The area consists of general buildings, and the visual context shows industrial/commercial structures that appear relatively robust. High population density (9,318.61 people per sq km) increases risk, but affluence and education suggest better infrastructure. Considering these factors, the damage is expected to be noticeable but not severe. \\
\hline
\textbf{Qwen-2.5-32B | 2014 Napa | Predicted MMI: VIII} \\
The earthquake has a magnitude of 6.0 Mw and occurred 10.24 km from the epicenter. The firm ground conditions (VS30 = 508 m/s) reduce shaking intensity. The area is sparsely populated (23.04 people per sq km), with low urbanization (2.97 \%) and high affluence (\$93,750 median income). The visual context shows a rural landscape with minimal infrastructure, indicating low vulnerability to structural damage. Based on these factors, the expected damage level is moderate. \\
\bottomrule[1.2pt]
\end{tabular}
\caption{Samples of LLM outputs with predicted MMI reasoning for selected earthquake scenarios.}
\label{tab:samples-llm-output}
\end{table*}

\end{document}